\documentclass[a4paper]{amsproc}

\usepackage{amssymb}
\usepackage{pstricks}
\usepackage{pstcol}
\usepackage{graphicx}
\usepackage{amsmath}
\usepackage{amsfonts}
\usepackage{latexsym}

 \advance\hoffset by -0.4 truecm

\newcommand{\R}{\text{\rm Re }}
\newcommand{\I}{\text{\rm Im }}
\def\F{\text{\bf F}}
\def\Vb{\text{\bf V}}

\begin{document}
\title[Generalizations of Kadanoff's exact solution...]
{Generalizations of Kadanoff's  solution of the Saffman--Taylor problem in a wedge}

\author[Irina Markina, Rodrigo Meneses, and Alexander Vasil'ev]{Irina Markina, Rodrigo Meneses, and Alexander Vasil'ev}

\address{\noindent Departamento de Matem\'atica,  Universidad T\'ecnica
Federico Santa Mar\'{\i}a,  Casilla  110-V,
Valpara\'{\i}so, Chile}

\thanks{This work is partially  supported by
Projects Fondecyt (Chile) \# 1030373, \#1040333, and UTFSM
\#12.05.23}

\email{\{irina markina, alexander.vasiliev\}@usm.cl}
\email{meneses.rod@gmail.com}

\subjclass[2000]{Primary: 76D27; Secondary: 30C35}

\keywords{Hele-Shaw problem, Saffman-Taylor finger, conformal map}

\begin{abstract}
We consider a zero-surface-tension two-dimensional Hele-Shaw flow in an infinite wedge. There exists a self-similar interface evolution in this wedge, an analogue of the famous Saffman-Taylor finger in a channel, exact shape of which has been given by Kadanoff. One of the main features of this evolution is its infinite time of existence and stability for the Hadamard ill-posed problem.
We derive several exact solutions existing infinitely  by generalizing and perturbing the one by Kadanoff.
\end{abstract}

\maketitle

\section{Introduction}
The Hele-Shaw problem involves two inmiscible Newtonian fluids that interact in a narrow gap between two parallel plates. One of them is of higher viscosity and the other is effectively inviscid. The model under consideration is valid when surface-tension effects in the plane of the cell are negligible. In the most of the cases it is known that when a fluid region is contracting, a finite time blow-up can occur, in which a cusp forms in the free surface. The solution does not exist beyond the time of blow-up. However, Saffman and Taylor in 1958  \cite{Saffman} discovered the long time existence of a continuum set of long bubbles within a receding fluid between two parallel walls in a  Hele-Shaw cell that further have been called the Saffman-Taylor fingers. It is worthy to mention that the first non-trivial explicit solution in the circular geometry has been
given by Polubarinova-Kochina and Galin in 1945 \cite{Galin, Polub1}. They also have proposed a complex variable approach, that nowadays is one of the principle tools to treat the Hele-Shaw problem in the plane geometry (see, e.g., \cite{How, Vas}) .  Following these first steps several other non-trivial exact solutions have been obtained (see, e.g.,
\cite{BA1, BA2, Cummings, Hoh, How4, Kadanoff, MarkVas}).  Through the similarity in the governing
equations (Hele-Shaw and Darcy), these solutions can be used to study  the models of saturated flows in porous media. Another typical scenario is given by Witten-Sander's diffusion-limited-aggregation (DLA) model (see, e.g., \cite{uu1}). In both cases the motion takes place in a Laplacian field (pressure for viscous fluid and random walker's probability of visit for DLA).
One of the ways, in which several new exact solution have been obtained, is to perturb known solutions. For example, Howison \cite{How2} suggested perturbations of the Saffman-Taylor fingers that led him to new fingering solutions keeping
the same asymptotic behavior as time $t\to \infty$. Recently, Hele-Shaw flows and Saffman-Taylor fingering phenomenon have been studied intensively in wedges (see, e.g., \cite{uu1, BA1, BA2, Cummings, Kadanoff, MarkVas} nad the references therein). In particular, Kadanoff \cite{Kadanoff} suggested a self-similar interface evolution between two  walls in a Hele-Shaw cell expressed explicitly by a rather simple parametric function  with a logarithmic singularity at one of the walls.
By this note we perturb Kadanoff's solution and give new explicit solutions with similar asymptotics.

\section{Mathematical model}

\begin{figure}[ht]
  \centering
\scalebox{1.0}{\includegraphics{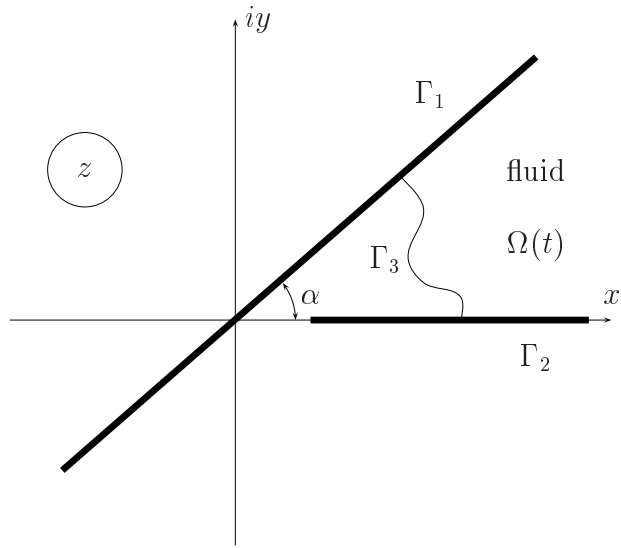}}
\caption[]{$\Omega(t)$
is the phase domain within an infinite corner and the homogeneous
sink/source at $\infty$}\label{fig3}
\end{figure}

We
suppose that the viscous fluid occupies a simply connected domain
$\Omega(t)$ in the phase $z$-plane whose boundary $\Gamma(t)$
consists of two walls $\Gamma_1(t)$ and $\Gamma_2(t)$ of the
corner and a free interface $\Gamma_3(t)$ between them at a moment
$t$. The inviscid fluid (or air) fills the complement to
$\Omega(t)$. The simplifying assumption of constant pressure at
the interface between the fluids means that we omit the effect of
surface tension. The velocity must be bounded close to the contact
point that yields the contact angle between the walls of the
wedge and the moving interface to be $\pi/2$ (see Figure
\ref{fig3}). A limiting case corresponds to one finite contact
point and the other tends to infinity. By a shift we can place the
point of the intersection of the wall extensions at the origin.
 To simplify
matter, we set the corner of angle $\alpha$ between the walls so
that the positive real axis $x$ contains one of the walls and fix
this angle as $\alpha\in (0,\pi]$.
In the zero-surface-tension model neglecting gravity, the unique acting force is
pressure $p(z,t)\equiv p(x,y,t)$. The velocity field averaged across the gap is given by the Hele-Shaw law (Darcy's law in
the multidimensional case) as $\Vb=-\nabla p$. Incompressibility implies that $p(z,t)$  is simply
\begin{equation}
\Delta p=0,\quad \mbox{in the flow region $\Omega(t)$.}\label{1}
\end{equation}
The dynamic condition
\begin{equation}
p\Big|_{\Gamma_3}=0,\label{2}
\end{equation}
is imposed on the free boundary $\Gamma_3\equiv\Gamma_3(t)$. The kinematic condition
implies that the normal velocity $v_n$ of the free boundary
$\Gamma_3$ outwards from $\Omega(t)$ is given as
\begin{equation}
 \frac{\partial p}{\partial
n}\Big|_{\Gamma_3}=-v_n.\label{3}
\end{equation}
On the walls $\Gamma_1\equiv\Gamma_1(t)$ and
$\Gamma_2\equiv\Gamma_2(t)$ the boundary conditions are given as
\begin{equation}
 \frac{\partial p}{\partial
n}\Big|_{\Gamma_1\cup \Gamma_2}=0, \label{4}
\end{equation}
(impermeability condition).
We suppose that the motion is driven by a homogeneous source/sink
at infinity. Since the angle between the walls at  infinity is
also $\alpha$, the pressure behaves about infinity as $$p\sim
\frac{-Q}{\alpha}\log |z|, \quad\mbox{as $|z|\to \infty$},$$ where
$Q$ corresponds to the constant strength of the source ($Q<0$) or
sink ($Q>0$). Finally, we assume that $\Gamma_3(0)$ is a given
analytic curve.

\begin{figure}[ht]
  \centering
\scalebox{1.0}{\includegraphics{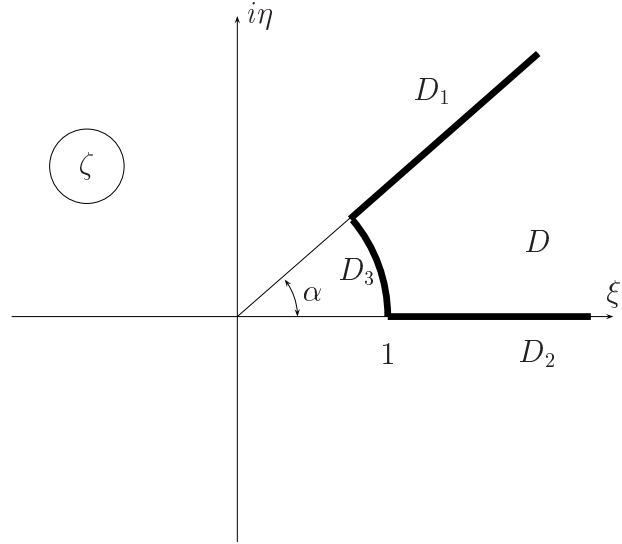}}
  \caption[]{The parametric domain $D$}\label{fig4}
\end{figure}

We introduce the complex velocity (complex analytic potential)
$W(z,t)=p(z,t)+i\psi(z,t)$, where $-\psi$ is the stream function.
Then, $\nabla p=\overline{\partial W/\partial z}$ by the
Cauchy-Riemann conditions. Let us consider an auxiliary parametric
complex $\zeta$-plane, $\zeta=\xi+i\eta$. We set
$D=\{\zeta:\,|\zeta|>1,\,0<\arg \zeta<\alpha\}$,
$D_3=\{z:\,z=e^{i\theta},\,\theta\in(0,\alpha)\}$,
$D_1=\{z:\,z=re^{i\alpha},\,r>1\}$, $D_2=\{z:\,z=r,\,r>1\}$,
$\partial D=D_1\cup D_2\cup D_3$, and construct a conformal
univalent time-dependent map $z=f(\zeta,t)$, $f:\,D\to \Omega(t)$,
so that being continued onto $\partial D$,
$f(\infty,t)\equiv\infty$, and the circular arc $D_3$ of $\partial
D$ is mapped onto $\Gamma_3$ (see Figure \ref{fig4}).
This map  has the expansion
$$f(\zeta,t)=\zeta\sum\limits_{n=0}^{\infty}a_{n}(t)\zeta^{-\frac{\pi
n}{\alpha}}$$ about infinity and $a_0(t)>0$. The function $f$
parameterizes the boundary of the domain $\Omega(t)$ by
$\Gamma_{j}=\{z:\,z=f(\zeta,t),\,\zeta\in D_{j}\}$, $j=1,2,3$.

We will use the notations $\dot f=\partial f/\partial t$,
$f'=\partial f/\partial \zeta$. The normal unit vector in the
outward direction is given by
$$\hat{n}=-\zeta\frac{f'}{|f'|}\,\,\,\mbox{on $\Gamma_3$, \ }
\hat{n}=-i \,\,\,\mbox{on $\Gamma_2$, and \ }  \hat{n}=i
e^{i\alpha}\,\,\,\mbox{on $\Gamma_1$.}$$ Therefore, the normal
velocity is obtained as
\begin{equation}
v_n=\Vb\cdot \hat{n}=-\frac{\partial p}{\partial n}= \left\{
\begin{array}{ll}
-\R\left(\displaystyle{\frac{\partial W}{\partial z}\frac{\zeta
f'}{|f'|}}\right), & \mbox{for $\zeta\in D_3$}\\ 0, & \mbox{for
$\zeta\in D_1$}\\ 0, & \mbox{for $\zeta\in D_2$}
\end{array}\right.\label{6}
\end{equation}
 The superposition $W\circ f$ is the solution to the mixed
boundary problem (\ref{1}), (\ref{2}), (\ref{4}) in $D$,
therefore, it is the Robin function given by $W\circ
f=-\frac{Q}{\alpha}\log \zeta$. On the other hand,
\begin{equation}
v_n= \left\{
\begin{array}{ll}
\R(\dot{f}{\overline{\zeta f'}}/{|f'|}), & \mbox{for $\zeta\in
D_3$}\\ \I(\dot{f}e^{-i\alpha}), & \mbox{for $\zeta\in D_1$}\\
-\I(\dot{f}), & \mbox{for $\zeta\in D_2$}
\end{array}\right.\label{7}
\end{equation}
The first lines of (\ref{6}), (\ref{7}) give us that
\begin{equation}
\R(\dot{f}\,\,{\overline{\zeta f'}})=\frac{Q}{\alpha},\quad
\mbox{for $\zeta\in D_3$}.\label{8}
\end{equation}
The resting lines of (\ref{6}), (\ref{7})  imply
\begin{equation}
\I(\dot{f}e^{-i\alpha})= 0\quad \mbox{for $\zeta\in D_1$,$\quad$ }
\I(\dot{f})= 0\quad \mbox{for $\zeta\in D_2$}. \label{9}
\end{equation}

\section{Exact solutions in a wedge of arbitrary angle}

We are looking for a solution in the
form
\[
f(\zeta,t)=\sqrt{\frac{2Qt}{\alpha}}\zeta+ \zeta g(\zeta),
\]
where $g(\zeta)$ is  regular in $D$ with the expansion
\[
g(\zeta)=\sum\limits_{n=0}^{\infty}\frac{a_n}{\zeta^{\frac{\pi n
}{\alpha}}}
\]
about infinity. The branch is chosen so that $g$, being continued
symmetrically into the reflection of $D$ is real at real points.
The equation (\ref{8}) implies that on $D_3$ the function $g$
satisfies the equation
\[
\R (g(\zeta)+\zeta g'(\zeta))=0,\quad \zeta\in D_3.
\]
Taking into account the expansion of $g$ we are looking for a
solution satisfying the equation
\begin{equation}
g(\zeta)+\zeta
g'(\zeta)=\frac{\zeta^{\frac{\pi}{\alpha}}-1}{\zeta^{\frac{\pi}{\alpha}}+1},\quad
\zeta\in D.\label{eq}
\end{equation}
Changing the right-hand side of the above equation one would
obtain other solutions. The general solution to (\ref{eq}) can be
given in terms of the Gauss hypergeometric function $\F\equiv
{_2\F}_1$ as
\[
\zeta
g(\zeta)=\zeta-2\zeta\F\left(\frac{\alpha}{\pi},1,1+\frac{\alpha}{\pi};
-\zeta^{\frac{\pi}{\alpha}}\right)+C.
\]
\begin{figure}{ht}
\centering
{\scalebox{1.0}{\includegraphics{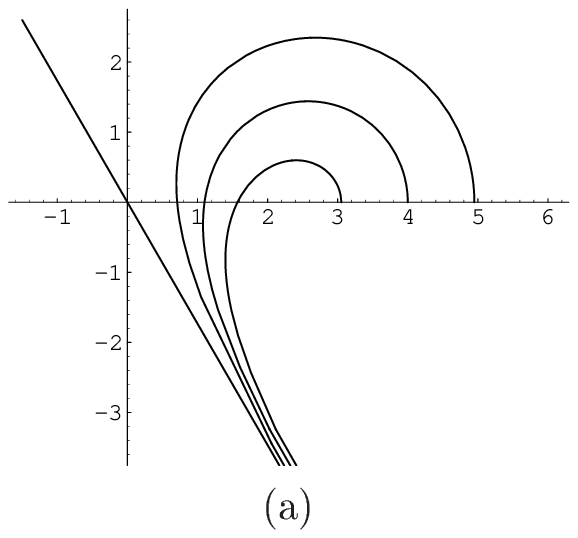}}}{\scalebox{1.0}{\includegraphics{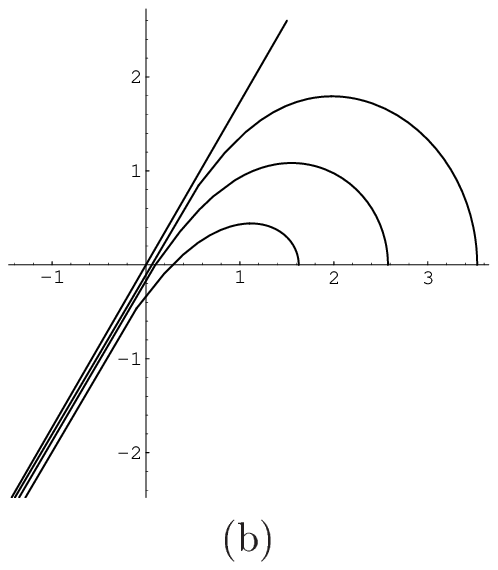}}}
\caption[]{Interface evolution in the wedge of angle: (a)
$\alpha=2\pi/3$; (b) $\alpha=\pi/3$}\label{fig14}
\end{figure}
We note that $f'$ vanishes for
$\zeta^{\frac{\pi}{\alpha}}=(2/(1+\sqrt{2Qt/\alpha}))-1$,
therefore, the function $f$ is locally univalent, the cusp problem
is degenerating and appears only at the initial time $t=0$ and the
solution exists during infinite time. The resulting function is
homeomorphic on the boundary $\partial D$, hence it is univalent
in $D$. This presents a case (apart from the trivial one) of the
long existence of the solution in the problem with suction
(ill-posed problem). To complete our solution we need to determine
the constant $C$. First of all we choose the branch of the
function $_2\F_1$ so that the points of the ray $\zeta>1$ have
real images. This implies that $\I C=0$. We continue verifying the
asymptotic properties of the function $f(e^{i\theta},t)$ as
$\theta\to\alpha-0$. The slope is
\[
\lim\limits_{\theta\to\alpha-0}\arg[ie^{i\theta}f'(e^{i\theta},t)]=
\alpha+\frac{\pi}{2}+\lim\limits_{\theta\to\alpha-0}\arg\left(\sqrt{\frac{2Qt}{\alpha}}+
\frac{e^{i\frac{\pi\theta}{\alpha}}-1}{e^{i\frac{\pi\theta}{\alpha}}+1}\right)=\alpha+\pi.
\]
To calculate shift we choose $C$ such that
\[
\lim\limits_{\theta\to\alpha-0}\I[e^{-i\alpha}f'(e^{i\theta},t)]=0.
\]
Using the properties of hypergeometric functions we have
\[
\lim\limits_{\gamma\to 0+0}\I
\F\left(\frac{\alpha}{\pi},1,1+\frac{\alpha}{\pi};
e^{i\gamma}\right)=\frac{\alpha}{2}.
\]
Therefore, $C=\alpha$. We present numerical simulation in
Figure~\ref{fig14}.

\begin{figure}[ht]
\centering
 \centerline{\scalebox{0.7}{\includegraphics{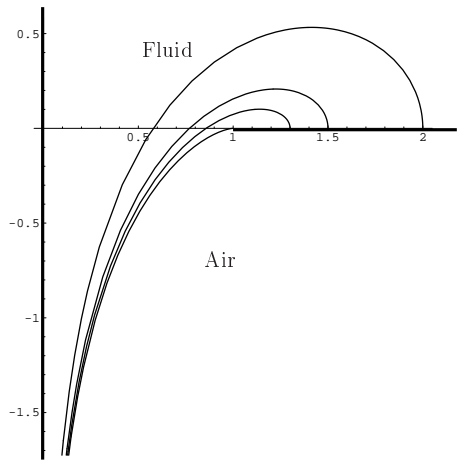}}}
\caption[]{Kadanoff's solution}\label{fig8}
\end{figure}

The special case of angle $\alpha=\pi/2$ has been considered by
Kadanoff \cite{Kadanoff}. The hypergeometric function is
reduced to  arctangent and we obtain
\begin{equation}
f(\zeta,t)=(\sqrt{4Qt/\pi}+1)\zeta+i\log\frac{1+i\zeta}{1-i\zeta}+\frac{\pi}{2},\quad
Q>0.\label{Kadanoff}
\end{equation}
 This function maps the domain
$\{|\zeta|>1,\,\,0<\arg\,\zeta<\pi/2\}$ onto an infinite domain
bounded by the imaginary axis $(\Gamma_1)$, the ray
$\Gamma_2=\{r:\,r\geq \sqrt{4Qt/\pi}+1\}$ of the real axis and an
analytic curve $\Gamma_3$ which is the image of the circular arc,
see Figure \ref{fig8}.

\section{Perturbations of Kadanoff's solution}

\begin{figure}[ht]
\centering
 \centerline{\scalebox{1.0}{\includegraphics{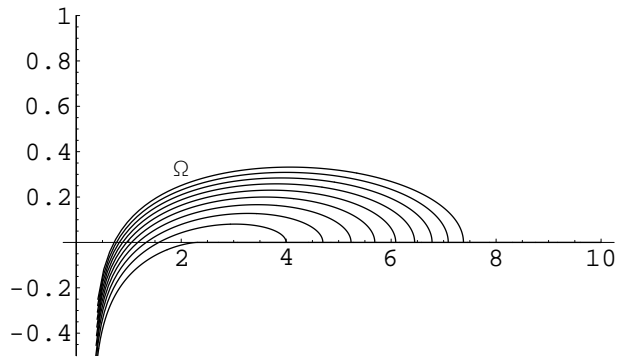}}}
\caption[]{Logarithmic perturbation of the elliptic evolution}\label{fig19}
\end{figure}

Kadanoff's solution (\ref{Kadanoff}) can be thought of as a logarithmic perturbation of a circular evolution
with the trivial solution $f_0(\zeta,t)=\sqrt{4Qt/\pi}\zeta$. A simple way to generalize the solution (\ref{Kadanoff})
is  to perturb another function. For example, one may choose
\[
f_0(\zeta,t)= A\sqrt{t}\left(c\zeta+\frac{1}{c\zeta}\right), \quad c>1, \quad A=\sqrt{\frac{4Qc^2}{\pi(c^4-1)}}.
\]
We find the solution $f(\zeta,t)$ in the form $f(\zeta,t)=f_0(\zeta,t)+ h(\zeta)$ similarly to the preceding
section. Then the equation (\ref{8}) is satisfied when
\[
\R \frac{\zeta h'(\zeta)}{\dot{f}_0(\zeta,t)}=0,\quad \mbox{or}\,\,\,\R\frac{\zeta h'(\zeta)}{c\zeta+1/c\zeta}=0,
\]
where $h'\sim (\zeta-i)^{-1}$ as $\zeta\to i$ in the unit circumference. We choose a consistent form
of $h$ as
\[
\frac{c\zeta^2 h'(\zeta)}{c^2\zeta^2+1}=\frac{\zeta^2-1}{\zeta^2+1}.
\]
Integration yields
\[
h(\zeta)=c\zeta+\frac{1}{c\zeta}-i\left(c-\frac{1}{c} \right)\log\frac{\zeta+i}{\zeta-i}+C,
\]
where $C$ is a constant of integration. Satisfying the conditions on the walls we deduce that $C=0$,
and finally, we get a logarithmic perturbation of the elliptic evolution as
\[
f(\zeta,t)=  (A\sqrt{t}+1)\left(c\zeta+\frac{1}{c\zeta}\right)- i\left(c-\frac{1}{c} \right)\log\frac{\zeta+i}{\zeta-i},
\]
see the interface evolution in
Figure~\ref{fig19}.

\begin{figure}[ht]
\centering
 \centerline{\scalebox{1.0}{\includegraphics{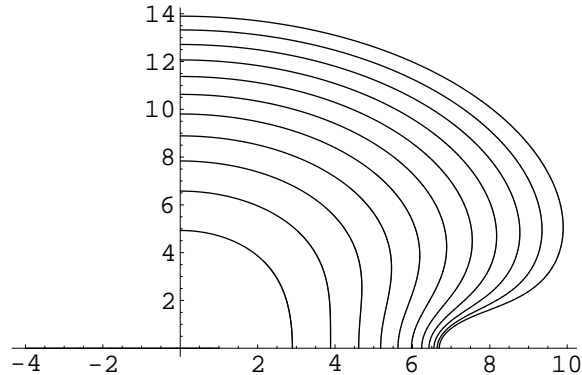}}}
\caption[]{Perturbation of Kadanoff's solution for $a(0)=0.5$, $d(0)=1$}\label{fig20}
\end{figure}

The next goal is to obtain perturbations of the logarithmic term of Kadanoff's solution that
with the same asymptotic as $t\to\infty$, such that the interface has finite contact points at a finite moment. Let us consider the function
\[
H(\zeta,t)=2d(t)\zeta-\log \frac{\zeta+a(t)}{\zeta-a(t)}.
\]
The functions $a(t)$, $d(t)$ are to be chosen such that  equation (\ref{8}) is satisfied for the moving interface as well as the conditions of impermeability and univalence hold. The local univalence
is followed from the first restriction $a(t)/d(t)<1$. Substituting $H$ into  equation (\ref{8}) and comparing the Fourier coefficients, we derive the following system of equations for the
functions $a(t)$ and $d(t)$:
\[
\left\{
\begin{array}{l}
(1+a^4)d\dot{d}- a^3\dot{d}+da^2\dot{a}-a\dot{a}=\frac{Q}{2\pi}(1+a^4),\\
   -2a^{2}d \dot{d}+a\dot{d}-\dot{a}d=-\frac{Q}{\pi}a^{2}
  \end{array}
   \right.
\]
This system can be easily solved and the first integrals are
\begin{equation}
d(t)=\frac{1+\sqrt{1+4a^2(t)(Qt/\pi-C_1)}}{2a(t)},\label{u1}
\end{equation}
\begin{equation}
2\frac{d(t)}{a(t)}-\log\frac{1+a^2(t)}{1-a^2(t)}=C_2,\label{u2}
\end{equation}
where
\[
 C_1=-d^2(0)+\frac{d(0)}{a(0)},\quad C_2=2\frac{d(0)}{a(0)}-\log\frac{1+a^2(0)}{1-a^2(0)},
\]
are the constants of integration. Let us assume the initial condition $a(0)\in (0,1)$. Making use of the system (\ref{u1},\ref{u2}) we arrive at the explicit
function $t(a)$ inverse to $a(t)$
\begin{equation}\label{in1}
t(a)=\frac{\pi}{Q}\left(\frac{\left(a^2\log\frac{1+a^2}{1-a^2}+a^2C_2-1\right)^2-1}{4a^2}+C_1\right),
\end{equation}
that exists, is continuous, and increases in the interval $a\in [a(0),1)$. Therefore, the function
$a(t)$ increases from $a(0)$ to $1$ as $t\in [0,\infty)$. By (\ref{u1}) we conclude that
$d(t)\sim O(\sqrt{t})$ as $t\to\infty$. The rotation of  $H$ is exactly Kadanoff's solution when
$a=1$, and $d(t)$ is appropriately chosen as in (\ref{Kadanoff}). To make a numerical simulation one may use the Newton method
of the solution of a non-linear system (Howison \cite{How2} presented the numerical approximation of an analogous solution in a narrow channel), see
Figure~\ref{fig20}.
Choosing $a(0)$ rather close to 1, one may give an explicit analytic approximation by, e.g., introducing two functions
\[
\hat{d}(t)=\frac{1+\sqrt{1+4(Qt/\pi-C_1)}}{2},
\]
\[
\hat{a}(t)=\sqrt{\frac{\exp(2\hat{d}(t)-C_2)-1}{\exp(2\hat{d}(t)-C_2)+1}}.
\]
The initial conditions $a(0)$ and $d(0)$ are to satisfy the inequalities $1-4C_1\geq 0$, $2\hat{d}(0)-C_2>0$. To proceed, we simplify
putting $a(0)=d(0)$. Then these inequalities are satisfied for $d(0)\in (\sqrt{3/4}, 1) $.
 It is easily seen from (\ref{u2}) that $|1-a(t)|\sim e^{-\sqrt{t}}$. Then $|\hat{d}(t)-d(t)|\sim e^{-\sqrt{t}}$ too.
Similarly, $\hat{d}(t)\sim \sqrt{t}$ and $|1-\hat{a}(t)|\sim e^{-\sqrt{t}}$. Both $\hat{a}(t)$ and $a(t)$
tend to 1 rapidly and the error  $|\hat{a}(t)-a(t)|$ is of the same order for $t\sim \infty$.

Now we evaluate the error $|\hat{a}(t)-a(t)|$ for $0<t<\infty$,
and claim that 
\begin{equation}\label{n1}0<a(t)-
\hat{a}(t)<8\big(a(0)- \hat{a}(0)\big).
\end{equation} To prove this
we estimate the distance $\rho(a)$ between the inverse
function~\eqref{in1} and $$\hat
t(a)=\frac{\pi}{Q}\left(\frac{\left(\log\frac{1+a^2}{1-a^2}+C_2-1\right)^2-1}{4}+C_1\right),
$$
as
$$\rho(a)=\hat t(a)-t(a)=(1-a^2)\left(\log\frac{1+a^2}{1-a^2}+C_2\right)^2\quad\text{and}\quad \rho(a(0))=2(1-a(0)).$$
The derivative of $\rho$ is
$$\rho^{\prime}(a)=2a\left(\log\frac{1+a^2}{1-a^2}+C_2\right)\left(\frac{4}{1+a^2}-\log\frac{1+a^2}{1-a^2}-C_2\right).$$
Since the function $\log\frac{1+a^2}{1-a^2}$ increases, the
function $\rho(a)$ may have a critical point $a_c$,
$a_c\in[a(0),1)$, which is the maximal solution to the
equation $ \frac{4}{1+a_c^2}=\log\frac{1+a_c^2}{1-a_c^2}-C_2$ in the interval $[a(0),1)$.
The latter equation implies
$$\rho(a_c)=(1-a_c^2)\left(\log\frac{1+a_c^2}{1-a_c^2}+C_2\right)^2
=\frac{16(1-a_c^2)}{\big(1+a_c^2\big)^2}\leq
16(1-a(0)^2)=8\rho(a(0)),$$ that proves~\eqref{n1}.

Moreover, $a(0)-\hat{a}(0)$ is decreasing and non-negative as a
function of the initial condition $a(0)$, that vanishes as
$a(0)\to 1$. Therefore, given a small positive number
$\varepsilon$, we may choose $a(0)=d(0)$ close to 1 such that
$\hat{a}(t)$ approximates $a(t)$ with the precision $\varepsilon$
during the whole time $0<t<\infty$. Desired quantity $a(0)$
satisfies the equation
$$a(0)-\sqrt{\frac{\exp{\Big(\sqrt{4a(0)^2-7}
+\log\frac{1+a(0)^2}{1-a(0)^2}-1\Big)}-1}{\exp{\Big(\sqrt{4a(0)^2-7}+\log\frac{1+a(0)^2}{1-a(0)^2}-1\Big)}+1}}
=\frac{\varepsilon}{8}.$$ A similar conclusion may be made for the function
$d(t)$ and its approximation $\hat{d}(t)$ (note that
$d(t)<\hat{d}(t)<1$). Moreover, the mapping
$$\hat{H}(\zeta,t)=2\hat{d}(t)\zeta-\log
\frac{\zeta+\hat{a}(t)}{\zeta-\hat{a}(t)}$$ converges to
Kadanoff's solution as $t\to\infty$.
\begin{figure}[ht]
\centering
 \centerline{\scalebox{1.0}{\includegraphics{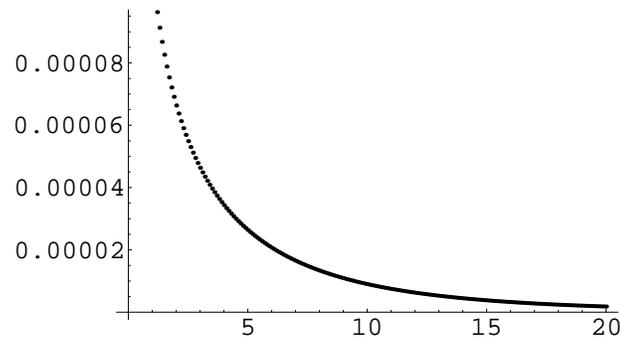}}}
\caption[]{Error  $|\hat{a}(t)-a(t)|$, for  $a(0)=0.9$, $d(0)=0.9$}\label{figerror}
\end{figure}

\end{document}